\documentclass[twocolumn,floatfix, showpacs]{revtex4}

\usepackage[final]{epsfig}
\usepackage{amsmath}
\usepackage{parskip}
 
\begin{document}
 
\title{Angular variation of hard back-to-back hadron suppression in heavy-ion collisions}
 
\author{Thorsten Renk}
\email{trenk@phys.jyu.fi}
\affiliation{Department of Physics, P.O. Box 35 FI-40014 University of Jyv\"askyl\"a, Finland}
\affiliation{Helsinki Institute of Physics, P.O. Box 64 FI-00014, University of Helsinki, Finland}
 
\preprint{HIP-2006-46/TH}

\begin{abstract}
The basic idea of jet tomography is to infer information about the density evolution of the medium created in heavy-ion (A-A) collisions by studying the suppression of hard probes in an A-A environment as compared to the baseline process known from p-p collisions. 
The suppression of back-to-back correlations in heavy-ion collisions allows, due to a different geometrical bias, a view into the medium which is qualitatively different from the one offered by single hadron suppression. 
A control parameter for the suppression corresponding to a systematic variation of in-medium pathlengths and density can be obtained by studying collisions at finite impact parameter $b$. A systematic variation of pathlength can then be introduced by studying the suppression pattern as a function of the angle $\phi$ with the reaction plane. Using a 3-d hydrodynamical evolution model for the medium and a Monte-Carlo (MC) model which has been shown to successfully reproduce the measured suppression of back-to-back correlations in central collisions of Au-Au at 200 AGeV, we compute the suppression as a function of $\phi$ for $b$ of 2.4 fm. 4.5 fm, 6.3 fm and 7.5 fm. Given that this involves variations in both control parameters $b$ and $\phi$ a comparison with data should eventually allow to place strong constraints on the combination of energy loss model and medium evolution model.
\end{abstract}
 
\maketitle

\section{Introduction}

The original hope of studying medium-induced modifications to hard processes in heavy-ion collisions has been to do jet tomography on the medium \cite{Jet1,Jet2,Jet3,Jet4,Jet5,Jet6}. The underlying idea is as follows: Hard processes take place simultaneously with soft processes responsible for the creation of soft bulk matter. This places partons emerging from hard vertices inside the medium, and interactions with the soft medium subsequently lead to an energy transfer from hard partons to the soft medium, resulting in a suppression of observed hard hadron yield. The strength of the interaction with the medium reflects the density of the medium, thus one should be able to gain information about the medium density from the observed strength of the suppression. The chief observable considered so far has been the nuclear suppression factor of single hadrons $R_{AA}$ (cf. e.g. \cite{PHENIX,STAR-data}) which is the measured yield in A-A collisions divided by the yield in p-p collisions multiplied with the number of binary scatterings (i.e. the default expectation if there would not be soft processes forming a medium in an A-A collision).

Unfortunately, there are a number of problems when one tries to carry out this program in practice. First, $R_{AA}$ is not a sensitive measure of the averaged energy loss probability distribution $\langle P(\Delta E, E)\rangle_{T_{AA}}$ \cite{gamma-h,inversion}. Assuming that eikonal propagation is a good approximation for hard partons, this quantity is the complete momentum space information about energy loss --- given that the vertex of origin of any observed hard hadron cannot be known, production point, direction and hence the length of propagation through the medium must be averaged probabilistically, leaving a momentum space probability  distribution that a parton with original energy $E$ lost the amount $\Delta E$ to the medium. The fact that $\langle P(\Delta E, E)\rangle_{T_{AA}}$ cannot be reliably determined from $R_{AA}$ is rather unfortunate, as knowledge of the energy loss probability distribution in momentum space is necessary before any mapping into position space (and hence any determination of density profiles) can be attempted.

The second obstacle is that there is not even qualitative agreement what the precise nature of in-medium energy loss is. In addition to medium-induced radiation \cite{Jet1,Jet2,Jet3,Jet4,Jet5,Jet6}, a number of publications have advocated a sizeable component of elastic energy loss \cite{Mustafa, Mustafa2, DuttMazumder,Djordjevic,Wicks,Qin} which has a qualitatively different pathlength dependence (linear in a homogeneous medium, as opposed to quadratic for radiative energy loss). As energy loss models typically contain one adjustable parameter linking the medium density with the interaction strength, $R_{AA}$ could be reproduced by a number of very different scenarios.

Some steps can be taken to improve the situation. First, instead of trying to determine an a priori unknown density evolution using hard probes, one may settle for the more modest goal of using a medium evolution model which is already constrained by soft hadronic observables and study the variation of the hard probes with changes of a known control parameter such as centrality. In this way, one employs hard probes as a test of a specific combination of evolution and energy loss model. Using the same 3-d hydrodynamical evolution which has been successful in describing bulk matter evolution \cite{Hydro3d}, it has been shown that different energy loss models predict a different suppression pattern as a function of reaction plane when going to off-central collisions even when all are adjusted to describe the observed suppression for central collisions \cite{RP1,RP2,RP3}.

One may also turn to observables in which the medium geometry is probed in a qualitatively different way, such as hard back-to-back correlations. Such correlations have been measured by the STAR collaboration \cite{STAR}. Employing again the strategy to use a set of a priori constrained medium evolution models, it could be shown in \cite{Correlations1,Correlations2} that the pathlength dependence of radiative energy loss is compatible with the strength of back-to-back suppression whereas a linear pathlength dependence characteristic of elastic energy loss is not \cite{ElasticPhenomenology}, thus any contribution with linear pathlength dependence must be a small correation.

In the present paper, we continue the work of \cite{RP1,Correlations2} in computing the suppression of back-to-back correlations for non-central collisions as a function of the angle with the reaction plane. Since this is a very differential observable, it constitutes a rather sensitive test of the radiative energy loss model \cite{Jet2,Jet4} in combination with the 3d-hydrodynamical evolution model \cite{Hydro3d}.

\section{The model}

The main ingredients of the model we wish to use have been extensively described in \cite{Hydro3d,RP1,Correlations1,Correlations2,dihadron-LHC}. We will therefore confine ourselves here to a summary and describe only the improvements made to the framework in more detail.

We calculate the correlation strength of hadrons back to back with a hard trigger in a MC simulation. There are three important building blocks to this computation: 1) the primary hard parton production, 2) the propagation of the partons through the medium and 3) the hadronization of the primary partons. Only step 2) probes properties of the medium, and hence it is here that we must specify details of the evolution of the medium and of the parton-medium interaction.

The strength of the parton-medium interaction contains one adjustable parameter $K$ (see below). This parameter is fixed by the requirement that the model should describe the suppression of single inclusive pions for central collisions. The results for non-central collisions, back-to-back correlations, different hadron species and different orientation with respect to the reaction plane are (given the hydrodynamical calculation as input) obtained without additional free parameters. 

\subsection{Primary parton production}

The production of two hard partons $k,l$ in leading order (LO) perturbative Quantum Choromdynamics (pQCD) is described by
 
\begin{equation}
\label{E-2Parton}
\frac{d\sigma^{AB\rightarrow kl +X}}{d p_T^2 dy_1 dy_2} \negthickspace = \sum_{ij} x_1 f_{i/A} 
(x_1, Q^2) x_2 f_{j/B} (x_2,Q^2) \frac{d\hat{\sigma}^{ij\rightarrow kl}}{d\hat{t}}
\end{equation}
 
where $A$ and $B$ stand for the colliding objects (protons or nuclei) and $y_{1(2)}$ is the 
rapidity of parton $k(l)$. The distribution function of a parton type $i$ in $A$ at a momentum 
fraction $x_1$ and a factorization scale $Q \sim p_T$ is $f_{i/A}(x_1, Q^2)$. The distribution 
functions are different for the free protons \cite{CTEQ1,CTEQ2} and nucleons in nuclei 
\cite{NPDF,EKS98}. The fractional momenta of the colliding partons $i$, $j$ are given by
$ x_{1,2} = \frac{p_T}{\sqrt{s}} \left(\exp[\pm y_1] + \exp[\pm y_2] \right)$.

Expressions for the pQCD subprocesses $\frac{d\hat{\sigma}^{ij\rightarrow kl}}{d\hat{t}}(\hat{s}, 
\hat{t},\hat{u})$ as a function of the parton Mandelstam variables $\hat{s}, \hat{t}$ and $\hat{u}$ 
can be found e.g. in \cite{pQCD-Xsec}. By selecting pairs of $k,l$ while summing over all allowed combinations of $i,j$, i.e. 
$gg, gq, g\overline{q}, qq, q\overline{q}, \overline{q}\overline{q}$ where $q$ stands for any of the quark flavours $u,d,s$
we find the relative strength of different combinations of outgoing partons as a function of $p_T$.

For the present investigation, we require $y_1 = y_2 = 0$, i.e. we consider only back-to-back correlations detected at midrapidity. In a first step, we sample Eq.~(\ref{E-2Parton}) summed over all $k,l$ to generate $p_T$ for the event, in the second step we perform a MC sampling of the decomposition of Eq.~(\ref{E-2Parton}) according all possible combinations of outgoing partons $k,l$ at the $p_T$ obtained in the first step. We thus end with a back-to-back parton pair with known parton types and flavours at transverse momentum $p_T$.

To account for various effects, including higher order pQCD radiation, transverse motion of partons in the nucleon (nuclear) wave function and effectively also the fact that hadronization is not a collinear process, we fold into the distribution an intrinsic transverse momentum $k_T$ with a Gaussian distribution, thus creating a momentum imbalance between the two partons as ${\bf {p_T}_1} + {\bf {p_T}_2} = {\bf k_T}$.

\subsection{Parton propagation through the medium}

The probability density $P(x_0, y_0)$ for finding a hard vertex at the 
transverse position ${\bf r_0} = (x_0,y_0)$ and impact 
parameter ${\bf b}$ is in leading order given by the product of the nuclear profile functions as
\begin{equation}
\label{E-Profile}
P(x_0,y_0) = \frac{T_{A}({\bf r_0 + b/2}) T_A(\bf r_0 - b/2)}{T_{AA}({\bf b})},
\end{equation}
where the thickness function is given in terms of Woods-Saxon the nuclear density
$\rho_{A}({\bf r},z)$ as $T_{A}({\bf r})=\int dz \rho_{A}({\bf r},z)$.  Note that Eq.~(\ref{E-Profile}) may receive (presumably) small corrections when going beyond a leading order calculation.

If we call the angle between outgoing parton and the reaction plane $\phi$, 
the path of a given parton through the medium $\xi(\tau)$ is specified 
by $({\bf r_0}, \phi)$ and we can compute the energy loss 
probability $P(\Delta E)_{path}$ for this path. We do this by 
evaluating the line integrals
\begin{equation}
\label{E-omega}
\omega_c({\bf r_0}, \phi) = \int_0^\infty \negthickspace d \xi \xi \hat{q}(\xi) \quad  \text{and} \quad \langle\hat{q}L\rangle ({\bf r_0}, \phi) = \int_0^\infty \negthickspace d \xi \hat{q}(\xi)
\end{equation}
along the path where we assume the relation
\begin{equation}
\label{E-qhat}
\hat{q}(\xi) = K \cdot 2 \cdot \epsilon^{3/4}(\xi) (\cosh \rho - \sinh \rho \cos\alpha)
\end{equation}
between the local transport coefficient $\hat{q}(\xi)$ (specifying 
the quenching power of the medium), the energy density $\epsilon$ and the local flow rapidity $\rho$ with angle $\alpha$ between flow and parton trajectory \cite{Flow1,Flow2}. Energy density $\epsilon$ and local flow rapidity $\rho$ are the input from the 3-d hydrodynamical simulation of the medium evolution \cite{Hydro3d}.

$\omega_c$ is the characteristic gluon frequency, setting the scale of the energy loss probability distribution, and $\langle \hat{q} L\rangle$ is a measure of the path-length weighted by the local quenching power.
We view  the parameter $K$ as a tool to account for the uncertainty in the selection of $\alpha_s$ and possible non-perturbative effects increasing the quenching power of the medium (see discussion in \cite{Correlations2}) and adjust it such that pionic $R_{AA}$ for central Au-Au collisions is described. This leads to a value of $K=3.6$ \cite{RP1}.

Using the numerical results of \cite{QuenchingWeights}, we obtain $P(\Delta E; \omega_c, R)_{path}$ 
for $\omega_c$ and $R=2\omega_c^2/\langle\hat{q}L\rangle$ for given jet production vertex and angle $\phi$.
In the MC simulation, we first sample Eq.~(\ref{E-Profile}) to determine the vertex of origin. For a given choice of $\phi$, we then propagate both partons through the medium evaluating Eqs.~(\ref{E-omega}) and use the output to determine $P(\Delta E; \omega_c, R)_{path}$ which we sample to determine the actual energy loss of both partons in the event.

\subsection{Hadronization}

\begin{figure*}[htb]
\epsfig{file=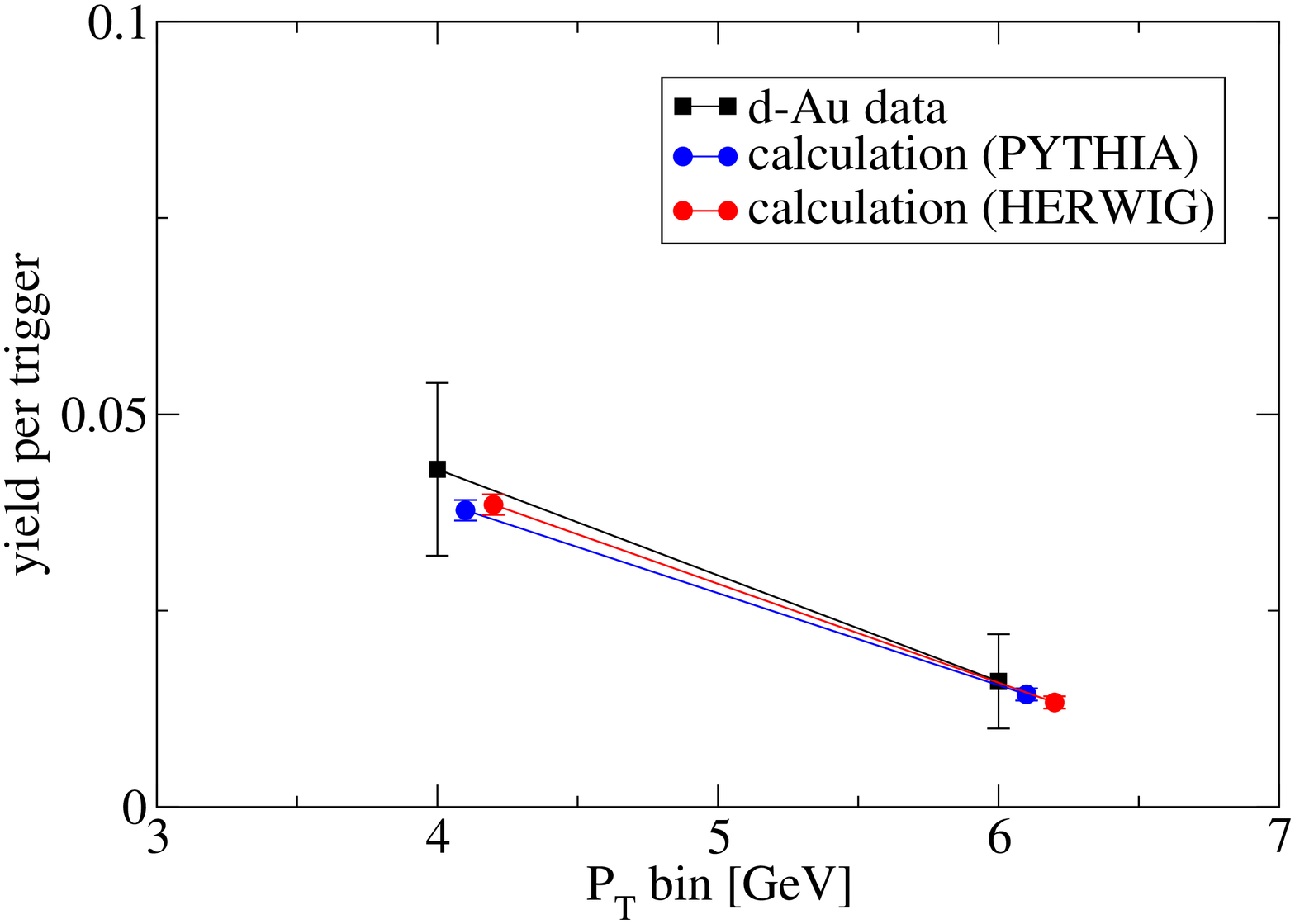, width=8.5cm} \epsfig{file=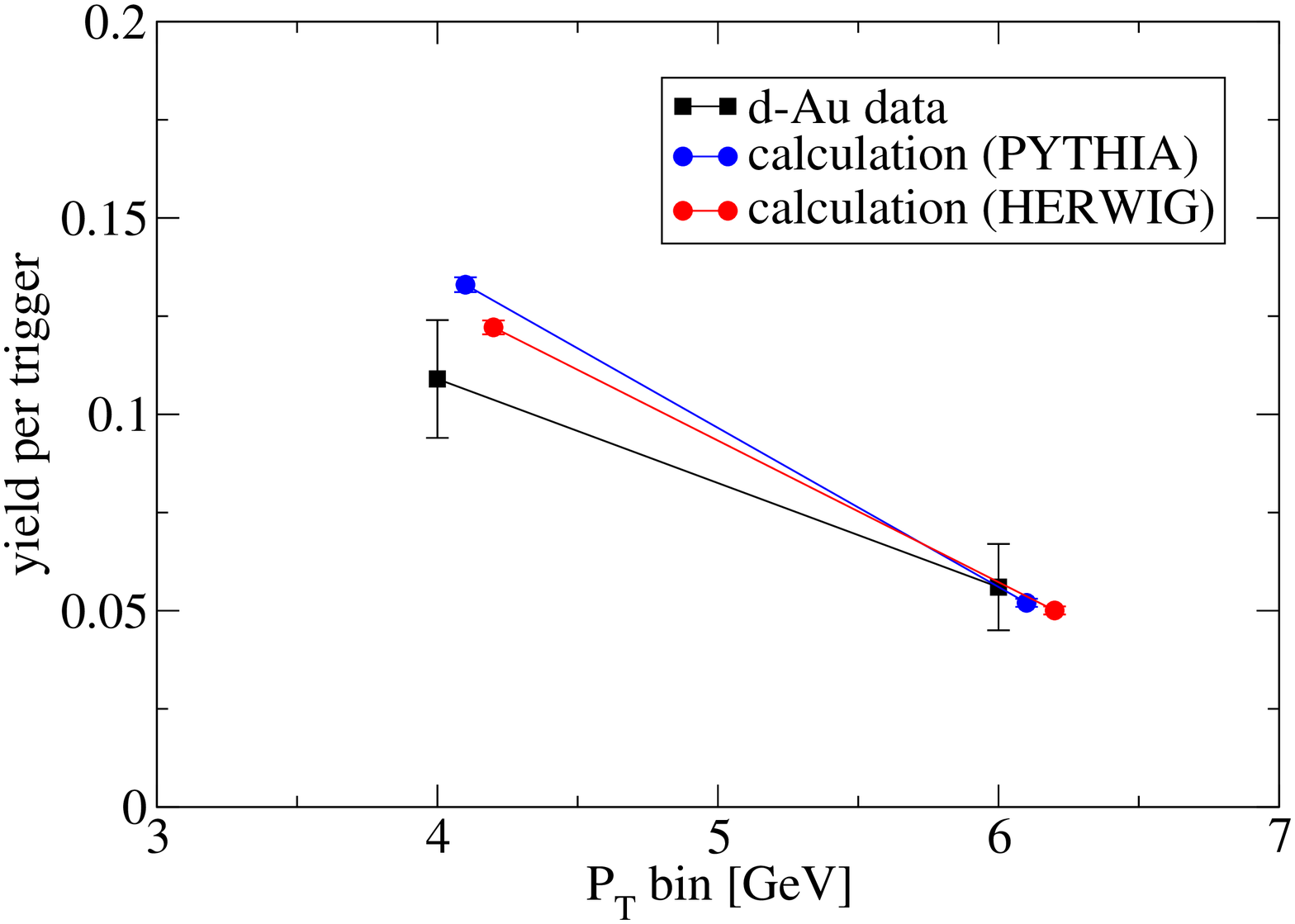, width=8.5cm} 
\caption{\label{F-Hadronization}(Color online) Comparison of the 200 AGeV d-Au baseline measurement \cite{STAR} for the near  (left panel) and away side (right panel) correlation strength with calculations utilizing PYTHIA \cite{PYTHIA} or HERWIG \cite{HERWIG} for shower development and hadronziation.}
\end{figure*}

Finally, we convert the simulated partons into hadrons. More precisely, in order to determine if there is a trigger hadron above a given threshold, given a parton $k$ with momentum $p_T$, we need to sample $A_1^{k\rightarrow h}(z_1, p_T)$, i.e. the probability distribution to find a hadron $h$ from the parton $k$ where $h$ is the most energetic hadron of the shower and carries the momentum $P_T = z_1 \cdot p_T$. In the following, we make the assumption that the hadronization process itself, at least for the leading hadrons of a shower, happens well outside the medium and as a consequence neglect any interaction of formed hadrons with the medium. The timescale for hadronization of a hadron $h$ in its restframe can be estimated by the inverse hadron mass, $\tau_h \sim 1/m_h$, boosting this expression to the lab frame one finds $\tau_h \sim E_h/m_h^2$. Inserting a hard scale of 6 GeV or more for the hadron energy and the pion mass in the denominator (as pions constitute the bulk of hadron production), this assumption seems well justified.

In previous works \cite{Correlations1,Correlations2} we have approximated this by the normalized fragmentation function $D_{k\rightarrow h}(z, P_T)$, sampled with a lower cutoff $z_{min}$ which is adjusted to the reference d-Au data. This procedure can be justified by noting that only one hadron with $z> 0.5$ can be produced in a shower, thus above $z=0.5$ the $D_{k\rightarrow h}(z, P_T)$ and $A_1^{k\rightarrow h}(z_1, p_T)$ are  (up to the scale evolution) identical, and only in the region of low $z$ where the fragmentation function describes the production of multiple hadrons do they differ significantly.

We improve on these results by extracting $A_1(z_1, p_T)$ and the conditional probability to find the second most energetic hadron at momentum fraction $z_2$ {\em given that the most energetic hadron was found with fraction $z_1$} $A_2(z_1, z_2, p_T)$ from a shower evolution code. The procedure is described in detail in \cite{dihadron-LHC} where we used PYTHIA \cite{PYTHIA} to simulate the shower, in the present paper we employ HERWIG \cite{HERWIG} instead which provides a slightly better description of the d-Au baseline data. A comparison of both approaches with the d-Au data is shown in Fig.~\ref{F-Hadronization}.

Our way of modelling hadronization corresponds to an expansion of the shower development in terms of a tower of conditional probability denities $A_N(z_1, \dots, z_n, \mu)$ with the probability to produce $n$ hadrons with momentum fractions $z_1, \dots z_n$ from a parton with momentum $p_T$  being $\Pi_{i=1}^n A_i(z_1,\dots z_i,p_T)$. Taking the first two terms of this expansion is justified as long as we are interested in sufficiently hard correlations.

Sampling $A_1(z_1, p_T)$ for any parton which emerged with sufficient energy from the medium provides the energy of the two most energetic hadrons on both sides of the event. The harder of these two is by definition the trigger hadron and defines the near side. The hadron opposite to it is then the leading contribution to the away side correlation.

The leading contribution to the correlation strength on the trigger side arises only in NL fragmentation, i.e. when sampling $A_2(z_1, z_2, p_T)$ for the (fixed) $z_1$ of the trigger to find the momentum of the second most energetic near side hadron. This term provides a correation to the away side correlation strength. This correction is $P_T$ dependent and decreases from 25\% in the lowest $P_T$ bin in the present investigation (4-6 GeV) to 2.7\% in the highest (10+ GeV) bin. Since we make a model comparison with the baseline both on near and away side, we compute both leading term and next-to-leading corection on the away side for consistency. 

\subsection{Hydrodynamic Description of the Medium}

For the hydrodynamical description of the soft medium, we use the 3-d relativistic fluid dynamics model by Bass and Nonaka \cite{Hydro3d}. The starting point is the
relativistic hydrodynamic equation
\begin{equation}
\partial_\mu T^{\mu \nu} = 0,
\label{Eq-rhydro}
\end{equation}
where $T^{\mu \nu}$ is the energy momentum tensor which is given by
\begin{equation}
T^{\mu \nu}=(\epsilon + p) U^{\mu} U^{\nu} - p g^{\mu \nu}.
\end{equation}
Here $\epsilon$, $p$, $U$ and $g^{\mu \nu}$ are energy density,
pressure, four velocity and metric tensor, respectively.
The relativistic hydrodynamic equation Eq.\ (\ref{Eq-rhydro})
is solved numerically using baryon number $n_B$ conservation
\begin{equation}
\partial_\mu (n_B (T,\mu) U^\mu)=0.
\end{equation}
as a constraint and closing the resulting set of partial
differential equations by specifying an equation of state (EoS):
$\epsilon = \epsilon(p)$.
In \cite{Hydro3d} a bag model EoS is used for the QGP.

The code
utilizes a Lagrangian mesh and 
light-cone coordinates $(\tau,x,y,\eta)$ ($\tau=\sqrt{t^2-z^2}$),
in order to optimize the model for ultra-relativistic regime
of heavy collisions at RHIC.
It is assumed that hydrodynamic expansion starts at
$\tau_0=0.6$ fm. Initial energy density and
baryon number density are parametrized by
\begin{eqnarray}
\epsilon(x,y,\eta)& =& \epsilon_{\rm max}W(x,y;b)H(\eta),
\nonumber \\
n_B(x,y,\eta)& = & n_{B{\rm max}}W(x,y;b)H(\eta),
\end{eqnarray}
where $b$ and  $\epsilon_{\rm max}$ ($n_{B{\rm max}}$) are
the impact parameter and the maximum value of energy density
(baryon number density), respectively.
$W(x,y;b)$ is given by a combination of wounded nuclear model and
binary collision model and  $H(\eta)$ is given
by $\displaystyle
H(\eta)=\exp \left [ - (|\eta|-\eta_0)^2/2 \sigma_\eta^2 \cdot
\theta ( |\eta| - \eta_0 ) \right ]$.
All parameters of the hydrodynamic evolution \cite{Hydro3d}
have been fixed
by a fit to the soft sector (elliptic flow, pseudo-rapidity
distributions and low-$p_T$ single particle spectra), therefore
providing a fully determined 
medium evolution for hard probes to propagate through.

Note that the fluid description in principle contains also hard hadrons from the high $P_T$ tails of the thermal distributions. However, these contributions are small, and while the soft medium may influence hadron production between 4 and 6 GeV due to recombiantion processes, hadron production above 6 GeV is clearly dominated by hard processes \cite{Coalescence, Coalescence2,Reco}. Consequently we neglect mechanisms of hadron production other than hard processes in the following.

\section{Results}

We compute the suppression of the away side correlation strength for a momentum range of 12--20 GeV for the trigger hadron. This choice is motivated by the results of \cite{Correlations2} where we found that information about the medium density becomes apparent only if the spread between trigger momentum and lowest associate momentum bin is large enough to allow not only particles with no energy loss to be registered but also include contributions from partons after finite energy loss. On the other hand, we found that the assumption that high $P_T$ hadron production can be modelled by jet fragmentation only holds for 6 GeV and above. Below, coalescence processes \cite{Coalescence,Coalescence2,Reco} play a significant role in hadron production in heavy-ion collisions. 

This requirement essentially determines the choice of the associate momentum windows: We include the window from 4--6 GeV for comparison, stressing that we expect that the model underpredicts the yield in this region. The remaining bins are chosen as 6--8 GeV, 8--10 GeV and 10+ GeV. In order to cancel systematic errors in the calculation of the baseline reaction without a medium, we present our results in the form of $I_{AA}$, i.e. the per-trigger yield in Au-Au collisions (for given $b$ and $\phi$) divided by the per-trigger yield in p-p collisions. 

Since we do not find any significant modifications on the near side (cf. the discussion in \cite{Correlations2}), we confine ourselves in the following to studying the modifications seen on the away side.

\subsection{$I_{AA}$ as a function of $P_T$, $b$ and $\phi$}

\begin{figure}[htb]
\epsfig{file=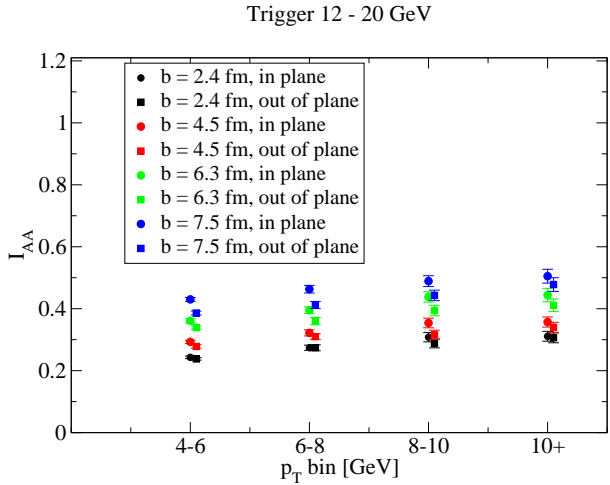, width=8cm}
\caption{\label{F-IAA-p_T}(Color online) $P_T$ dependence of charged hadron $I_{AA}$ in 200 AGeV Au-Au collisions for a 12-20 GeV trigger as a function of impact parameter $b$, shown for in-plane and out-of-plane emission.}
\end{figure}

We show the resulting $I_{AA}$ as a function of associate hadron momentum bin in Fig.~\ref{F-IAA-p_T} for four different values of impact parameter $b$ both for in-plane ($\phi=0$) and out-of-plane ($\phi = \pi/2$) emission.

The first feature observed is that there is no strong $P_T$ dependence observed in $I_{AA}$ for any impact parameter, albeit a small rise may be present. The strongest trend seen is that $I_{AA}$ shows a rise with $b$, corresponding to the fact that both entropy production (and hence medium density) and average pathlength are reduced for peripheral collisions as compared to central collisions. A growing split between in-plane and out-of-plane emission as a function of $b$ shows that the spatial asymmetry in the initial state of the hydrodynamical calculation is indeed translated into a $\phi$ dependent suppression of the correlation strength.

\begin{figure}[htb]
\epsfig{file=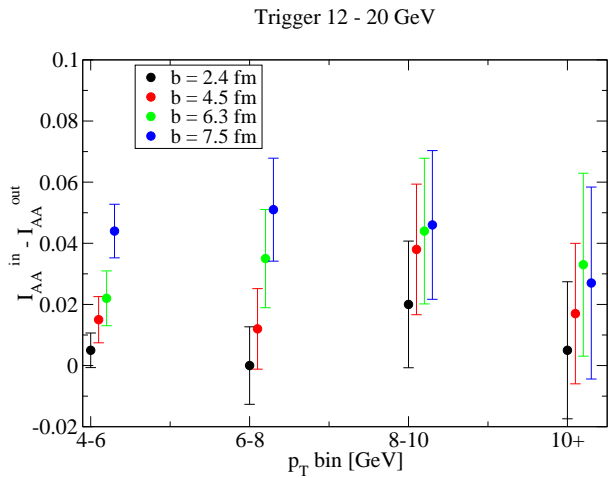, width=8cm}
\caption{\label{F-spread-p_T}(Color online) $P_T$ dependence of the spread between in-plane and out-of-plane emission in 200 AGeV Au-Au collisions for a 12-20 GeV trigger as a function of impact parameter $b$.}
\end{figure}

We show the $P_T$ dependence of the difference between in-plane and out-of-plane emission in Fig.~\ref{F-spread-p_T}. There is some indication that the separation between the different angular orientations is more pronounced at lower $P_T$, although the significant reduction of MC statistics in the high $P_T$ bins as compared to the low $P_T$ bins poses difficulties to a solid statement.

Surprisingly, the magnitude of both $I_{AA}$ and of the split between in-plane and out-of-plane emission as a function of $b$ is rather similar to the values found for single hadron suppression in terms of $R_{AA}$ found within the same formalism \cite{RP1}. However, $R_{AA}$ and $I_{AA}$ are in fact not the same, the similarity holds only on the level of about 30\%. For this particular set of trigger and assoicate momenta, the magnitude of $I_{AA}$ is larger than $R_{AA}$ whereas the spread between in-plane and out-of-plane emission is consistently smaller than the spread found in $R_{AA}$.

\subsection{The geometry of single hadron and dihadron suppression}

\begin{figure*}[htb]
\epsfig{file=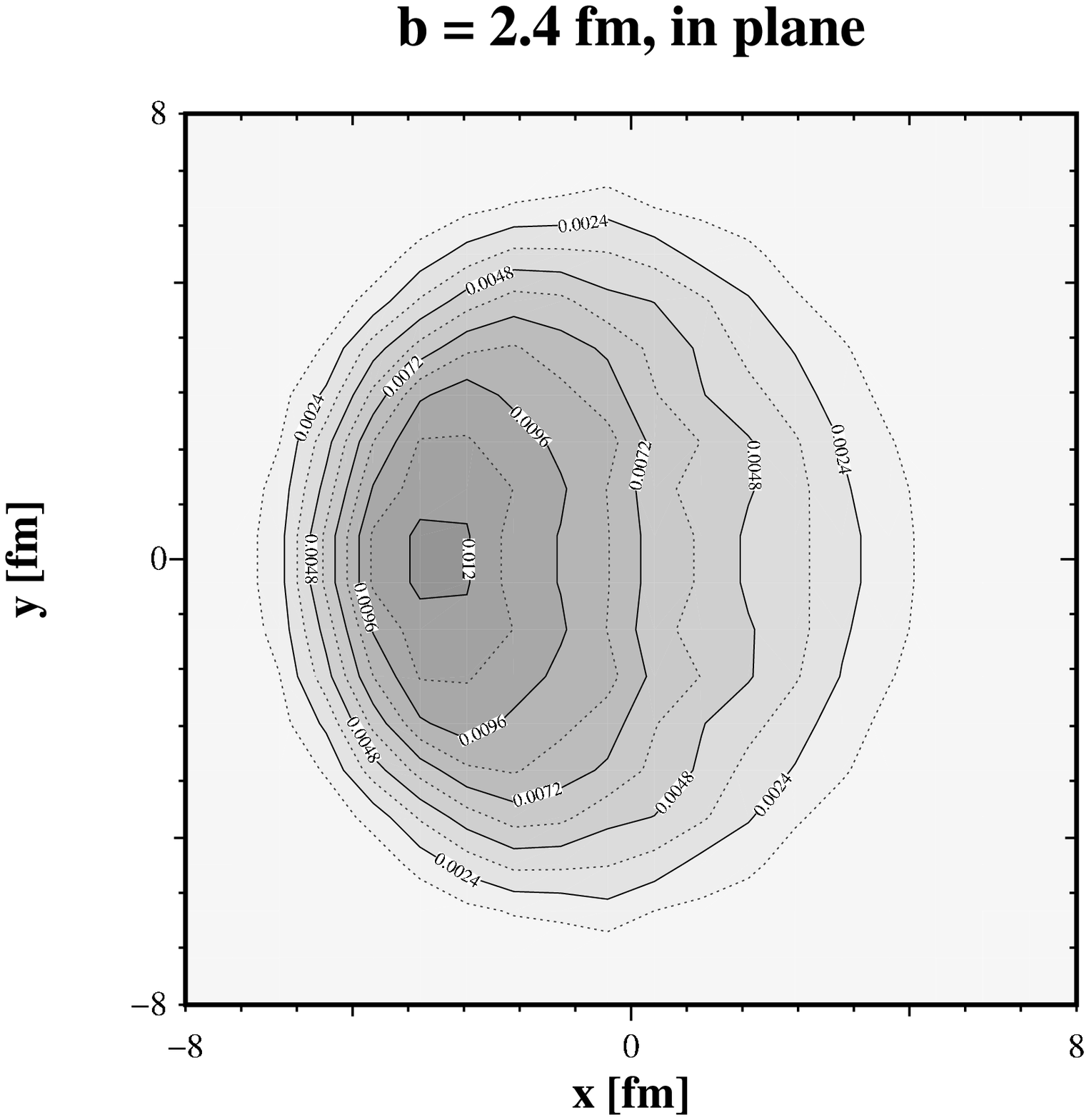, width=8cm} \epsfig{file=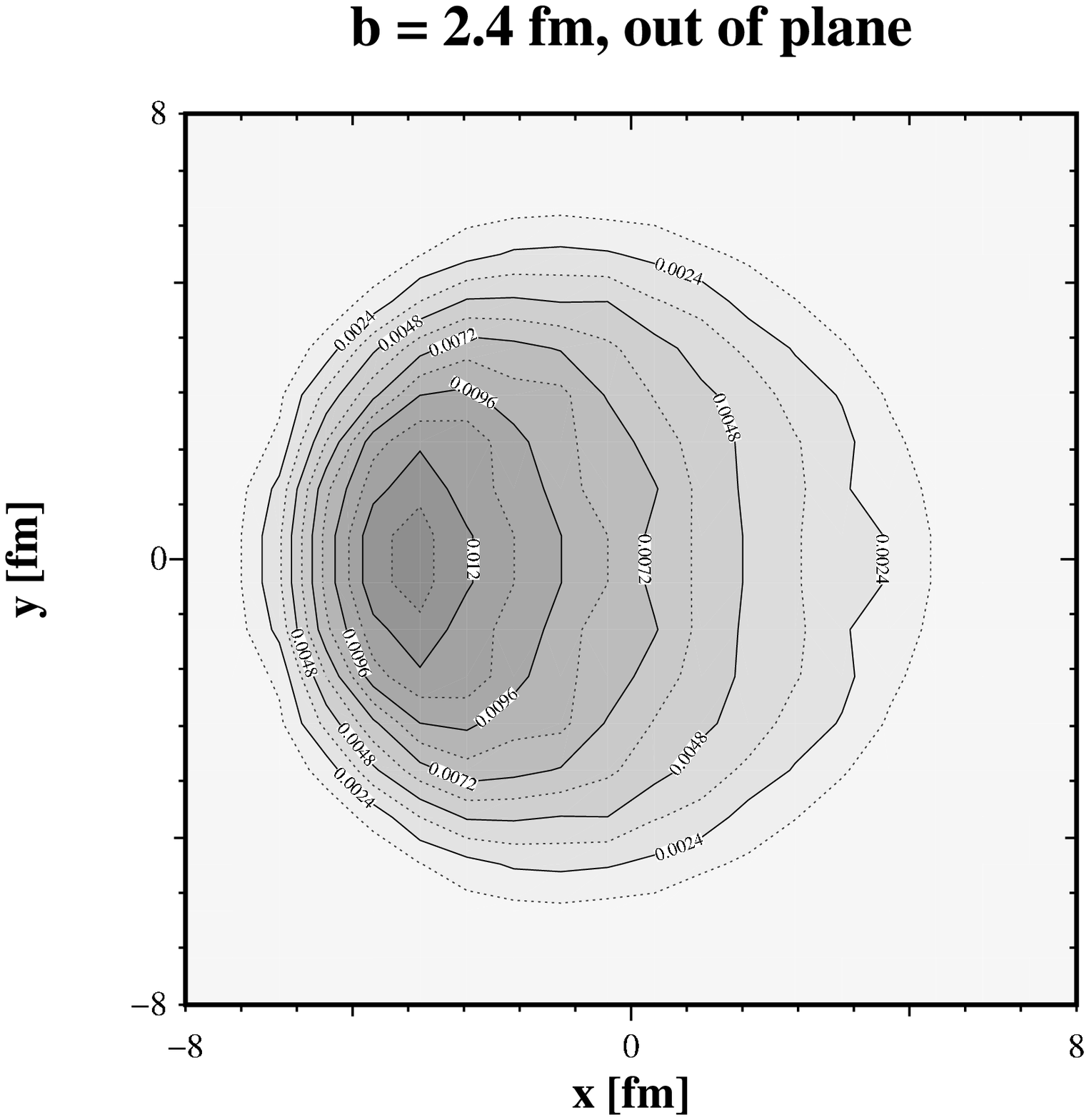, width=8cm}\\
\epsfig{file=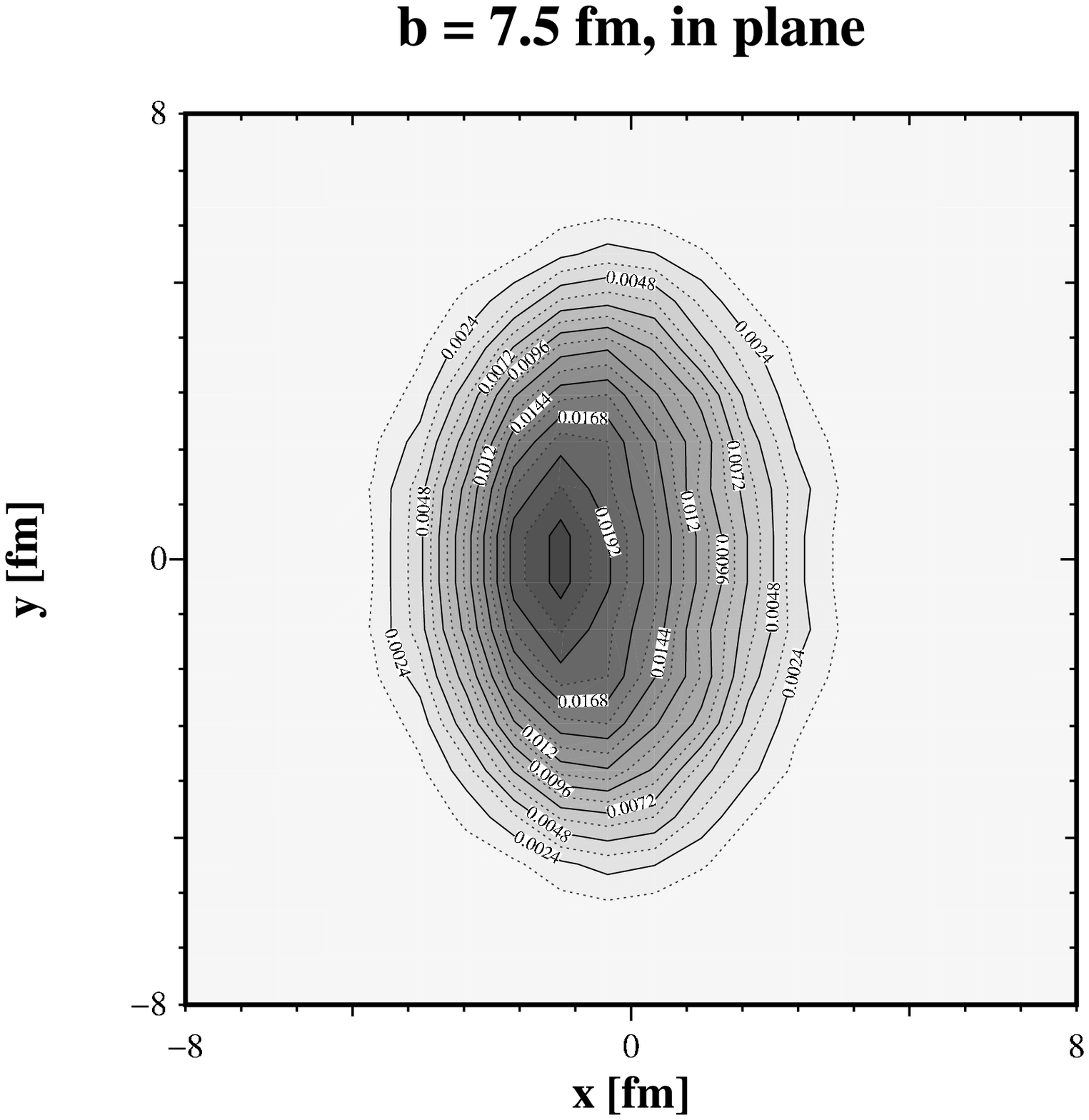, width=8cm} \epsfig{file=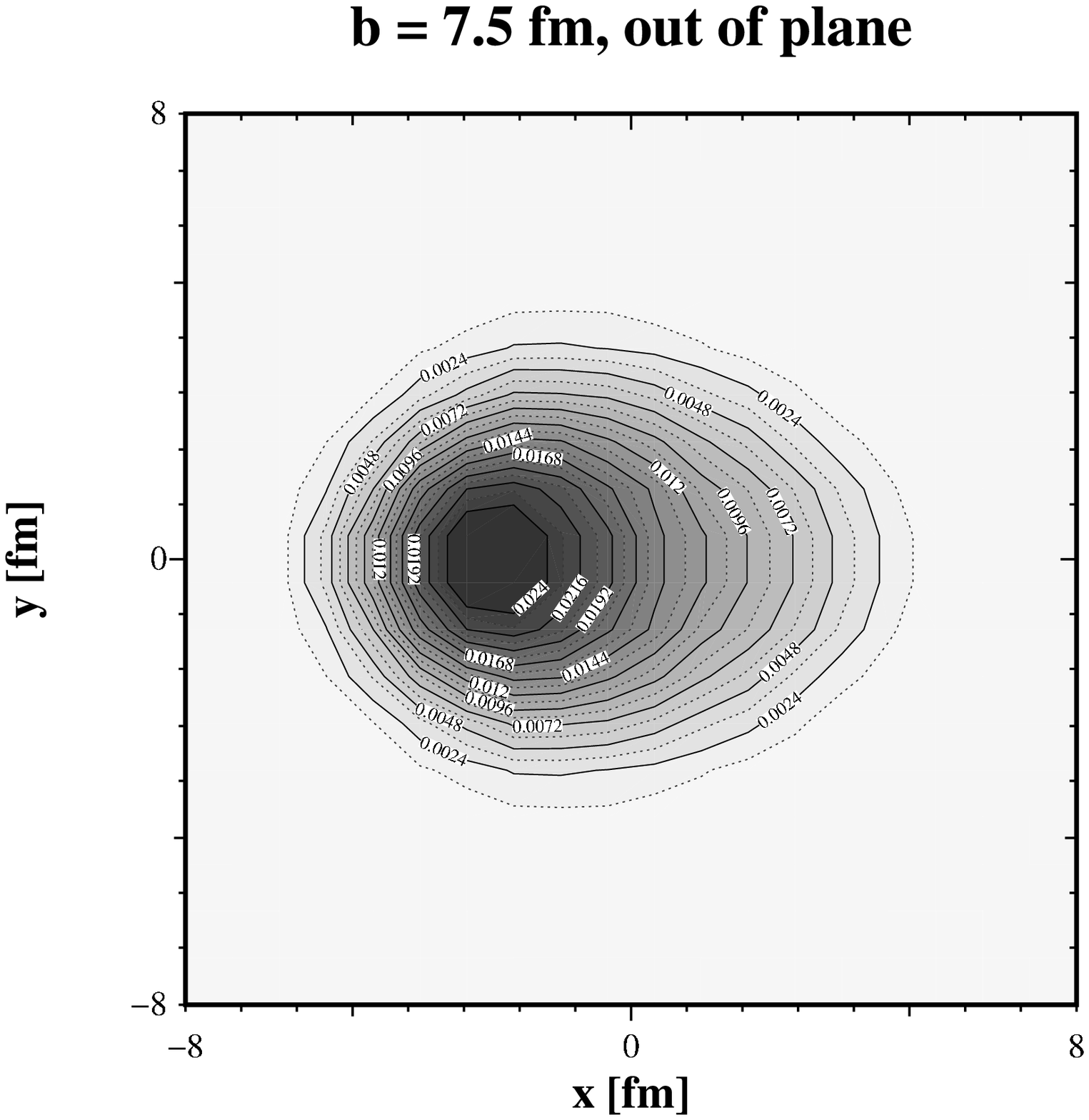, width=8cm}\\
\caption{\label{F-vdist}Probability density for an inclusive hard hadron measurement, i.e. for  finding a parton production vertex at $(x,y)$ given a triggered event with 12 GeV $< p_T <$ 20 GeV for different $b$ and $\phi$. In all cases the near side (triggered ) hadron propagates to the $-x$ direction. Countours are at linear intervals. }
\end{figure*}

We show the probability density  of vertices in the $(x,y)$ plane  leading to a near side trigger  hadron between 12 and 20 GeV in Fig.~\ref{F-vdist} for impact parameters $b = 2.4$ fm and $b = 7.5$ fm both for in-plane and out-of-plane emission. 

For the most central impact parameter $b=2.4$ fm the results of the simulation in the 3-d hydrodynamical medium evolution are very similar to the results obtained in \cite{Correlations2} for central Au-Au collisions in a 2-d hydrodynamics simulation \cite{Hydro}. There is a clear surface effect visible, the most likely point of origin of a triggered hadron is not in the medium center but about 4 fm displaced and hence close to the surface. The detailed position of this maximum arises from a balance between the distribution of primary vertices coming from the nuclear overlap Eq.~(\ref{E-Profile}) which peaks in the medium center and the fact that partons originating in a high-density region are more likely to be quenched. On the other hand, close to the medium edge where the quenching is small, the probability to find a vertex is much reduced. 

For $b=2.4$ fm, there is no large difference between in-plane and out-of-plane visible, corresponding to the fact that the spatial asymmetry of the medium is small. This is different for the simulation with $b=7.5$ fm. Here, clear changes in geometry when going from in-plane to out-of plane emission are visible. In particular, the surface effect is stronger for out-of-plane emission due to the greater amount of medium in the path of outgoing partons for this configuration.

Note that the distribution shown here is qualitatively (i.e. up to small changes with $P_T$) the same for any hard single hadron observable, in particular $R_{AA}$ (averaged over the trigger range) and $\gamma$-hadron correlations (where the photon escapes into the $+x$ direction).

\begin{figure*}[htb]
\epsfig{file=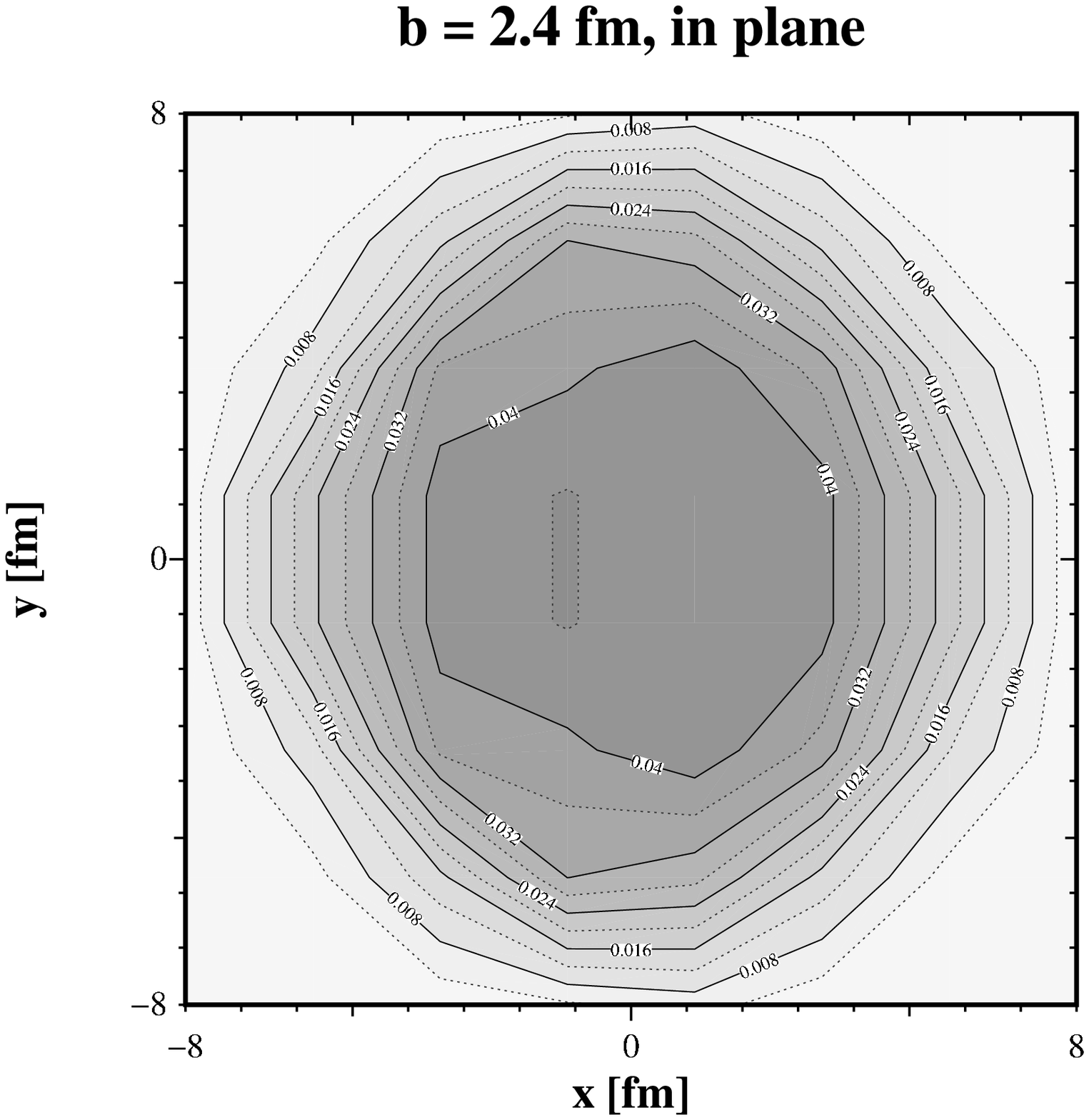, width=8cm} \epsfig{file=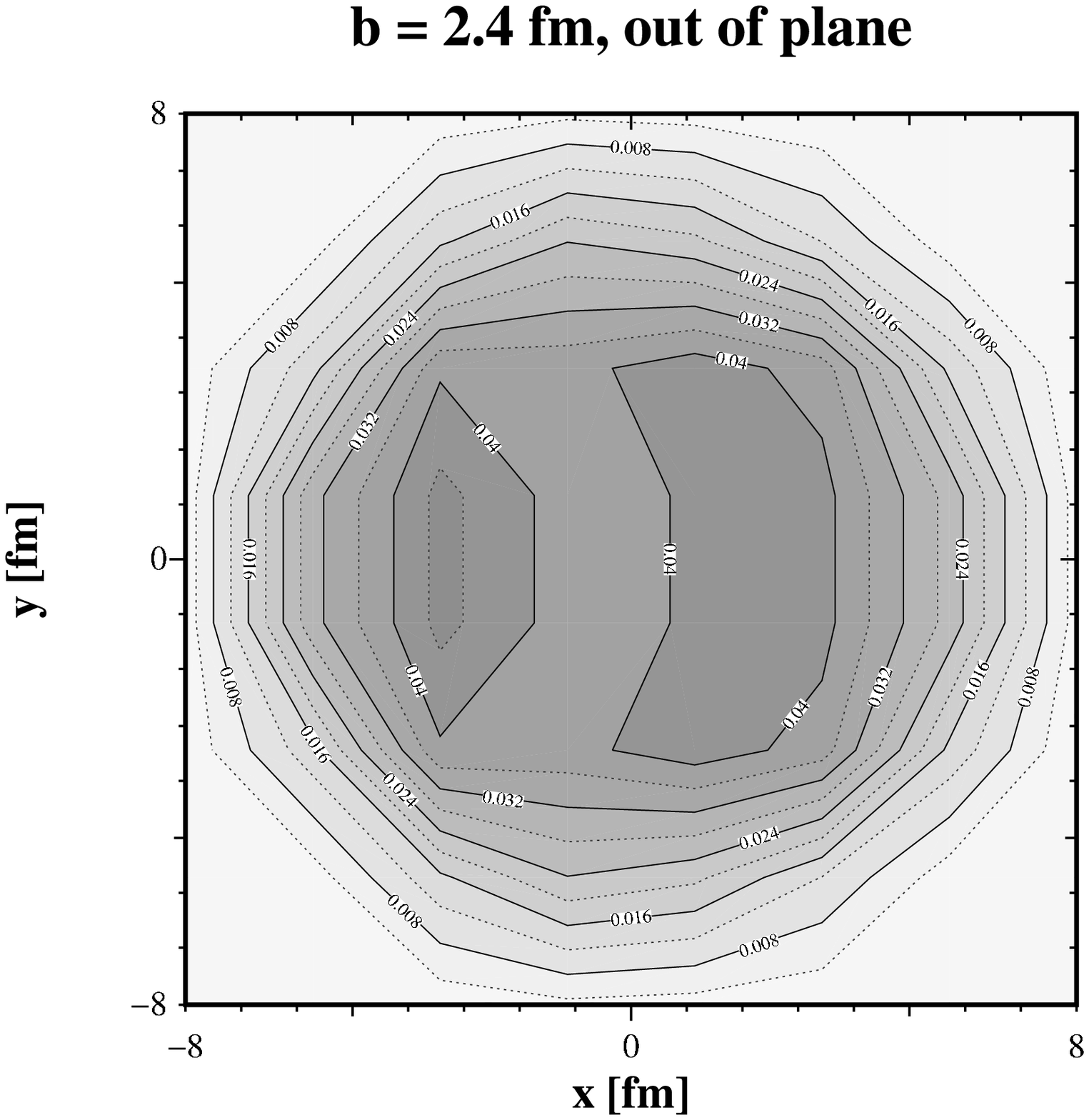, width=8cm}\\
\epsfig{file=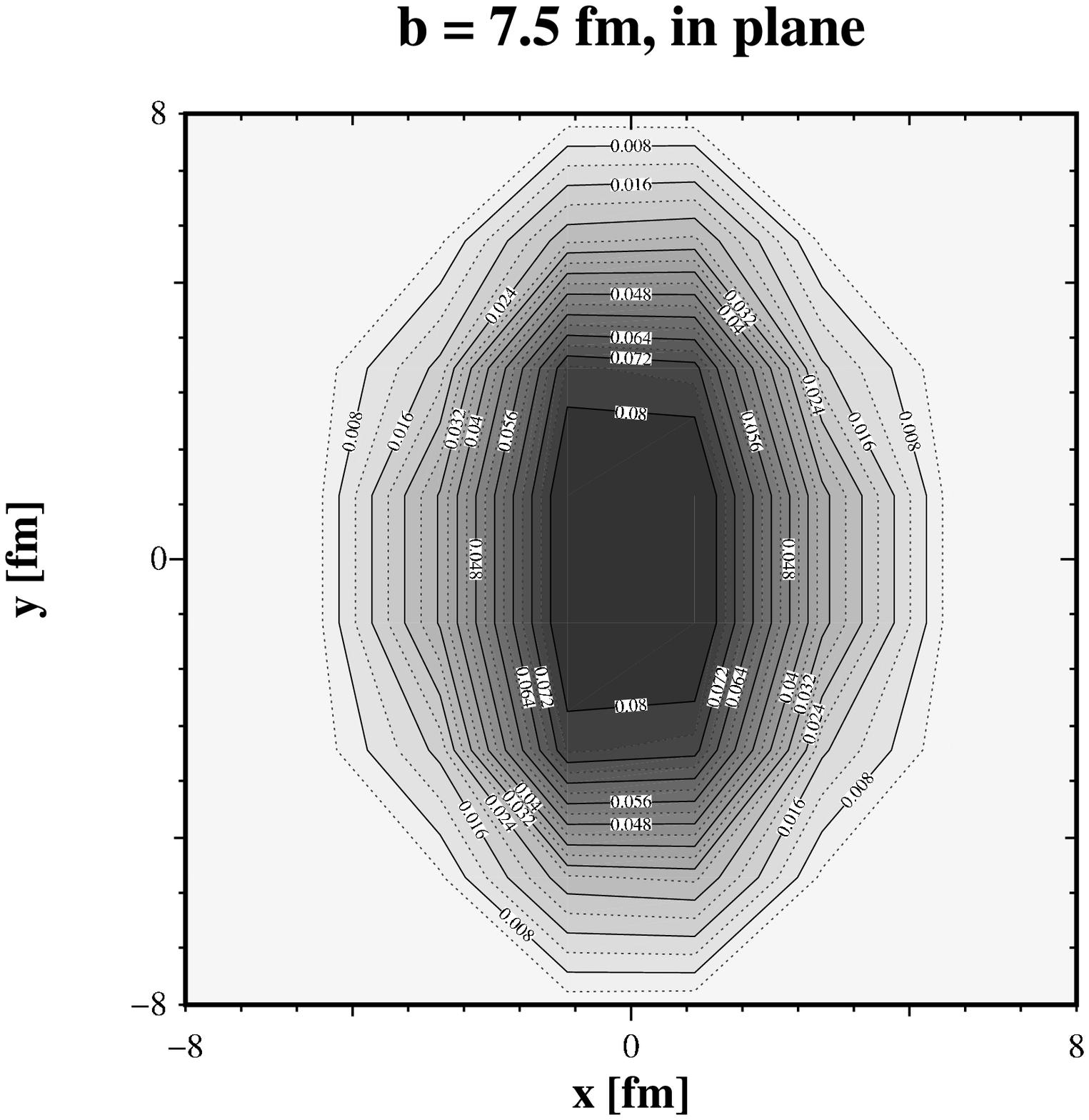, width=8cm} \epsfig{file=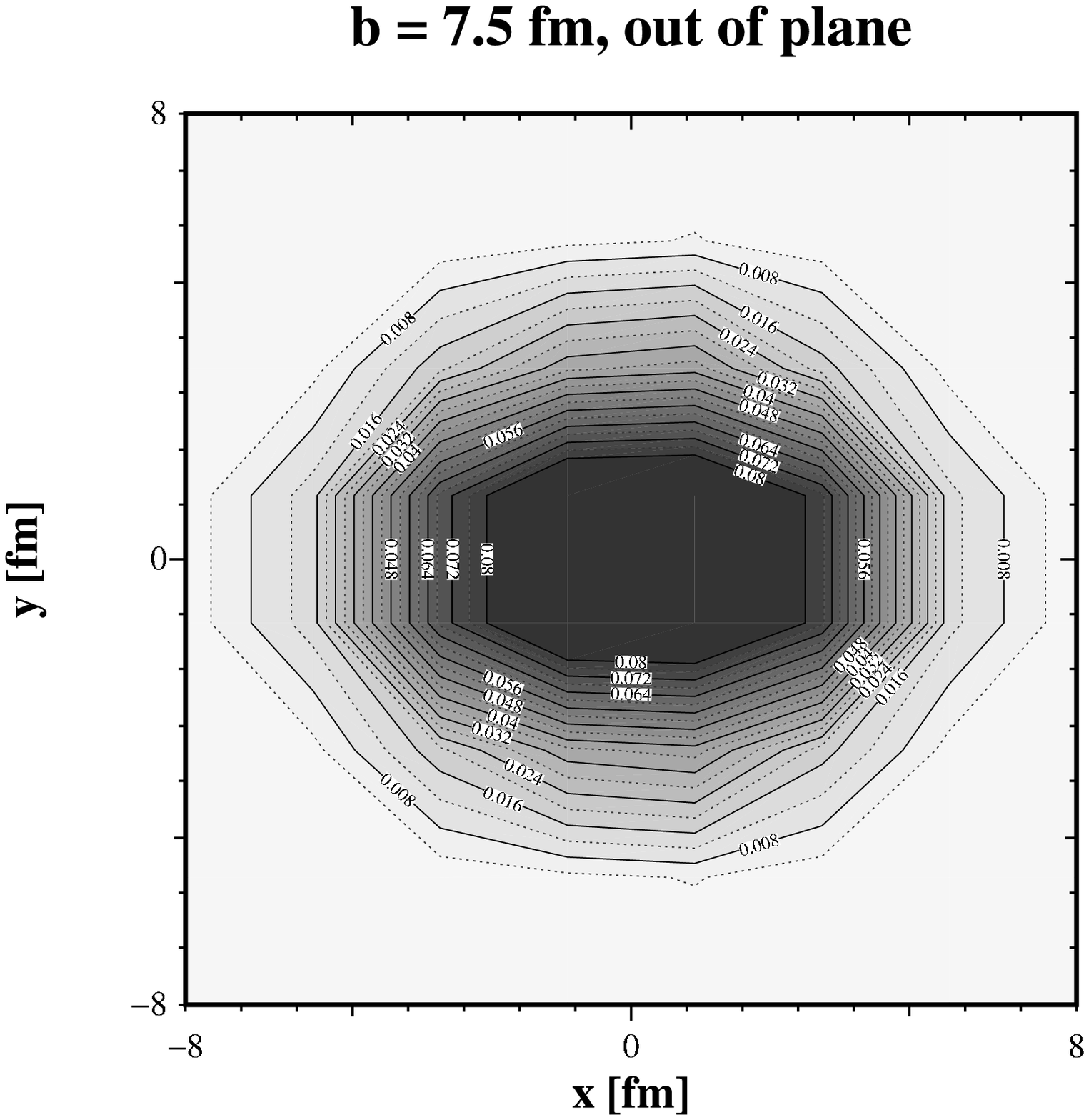, width=8cm}\\
\caption{\label{F-pdist}Probability density for a coincidence hard hadron measurement, i.e. for finding a vertex at $(x,y)$ leading to a triggered event with 12 GeV $< p_T <$ 20 GeV and in addition an away side hadron with 4 GeV $< p_T < 
6$ GeV for different $b$ and $\phi$. In all cases the near side hadron 
propagates to the $-x$ direction. Countours are at linear intervals.}
\end{figure*}

We show the probability density  of vertices in the $(x,y)$ plane  leading to a near side trigger 
hadron between 12 and 20 GeV and an associate away side hadron with 4 GeV $< p_T <$ 6 GeV in 
Fig.~\ref{F-pdist}. It is immediately obvious that the distribution is very different from the 
distribution of vertices for single hadron observation shown in Fig.~\ref{F-vdist}. First, the 
dihadron distributions are much wider in $\pm y$ direction, indicating the importance of the 
periphery where both near and away side parton have a short in-medium path (or the halo where the 
production vertex lies outside the medium). Second, the distribution is almost symmetric with respect to the
$y$-axis, indicating the events are favoured in which the near and away side pathlength is about equal.
Both features are consistent with the distributions observed in \cite{Correlations2} for different density evolution models.

In particular, note the absence of tangential emission, which in the framework presented here does occur, albeit at much stronger quenching power of the medium \cite{Correlations2}.

It is evident from the underlying geometry that the physics (in terms of average pathlength and relevant medium density) of single hadron and dihadron suppression is rather different in the calculation. Geometry alone clearly cannot account for the observed similarity between $R_{AA}$ and $I_{AA}$.

\section{Discussion}

One may be tempted to conclude from the similarity of $R_{AA}$ and $I_{AA}$ that there is no additional information contained in $I_{AA}$ beyond what can be obtained from single hadron suppression. However, this is clearly not the case. Let us illustrate in a qualitative way the meaning of this observation.

In a purely geometrical suppression picture, a parton is absorbed when the hard vertex of its origin is in a dense region but unmodified when the vertex falls into the dilute halo region. In this case, the whole back-to-back event is either absorbed or unmodified. This in turn implies that the number of triggered events is reduced, but once an event is triggered, there is no modification of the near or away side parton. From this one may immediately deduce that in this scenario $R_{AA}$ may be below 1, but $I_{AA}$ will always be (modulo small effects by the nuclear parton distribution functions) equal to 1. 

Thus, in any situation in which quenching is chiefly driven by the density at the vertex of origin and the medium can be modelled as an opaque and a dilute region, $I_{AA}$ is expected to be substantially larger than $R_{AA}$. Within this model, this trend has been demonstrated as a rise in the per-trigger yield in the black core scenario described in \cite{Correlations2}.

Let us next consider the momentum space. By definition, the trigger hadron is the hardest hadron of the event. Consequently, there is the strongest bias on the trigger parton to experience no or only a small amount of energy loss. In contrast, the bias {\em given that there is a hard hadron on the near side} on the away side hadron is considerable reduced, allowing average energy losses of several GeV on the away side \cite{Edep}. If the geometry-averaged energy loss probability on near and away side would be equal (i.e. if there would not be a bias towards surface emission), this would imply again that $I_{AA}$ should be substantially larger than $R_{AA}$ (for the measured $R_{AA}$ of about 0.2, this would amount to almost a factor 2 difference).

However, there is an opposing bias in position space: The fact that the hardest hadron tends to be produced close to the surface means that the away side parton on average has a longer pathlength. Here, the weighting of the energy loss with pathlength is crucial, as the medium density drops due to the longitudinal and radial expansion of the medium while the parton traverses the medium. An energy loss which grows linear in pathlength $L$ in a constant medium (as characteristic of elastic energy loss) for example  translates into a logarithmic pathlength dependence $\sim \ln(L/L_0)$ in a medium undergoing longitudinal Bjorken expansion where the density drops like $1/\tau$. In this situation, pathlength differences on near and away side do not matter much and the bias in momentum space is expected to win out. This is observed within the model described here in \cite{ElasticPhenomenology}.

On the other hand, for an energy loss with quadratic pathlength dependence $\sim L^2$ in a constant medium, the energy loss gets proportional to $\sim L$ in a medium undergoing longitudinal Bjorken expansion. If the away side path is on average twice as long as the near side path this effect would be expected to cancel the reduced bias in momentum space and make $R_{AA}$ and $I_{AA}$ about equal. This argument strongly relies on modelling the expansion of the medium correctly --- note that in the static medium studied in \cite{PQM} in a conceptually similar framework using the same radiative energy loss pciture, back-to-back events are almost completely suppressed, corresponding to an energy loss dependence with pathlength which is too strong.

These arguments may help to understand why the fact that $I_{AA}$ and $R_{AA}$ are of similar magnitude is in fact highly non-trivial and in itself probes properties of the medium and its expansion along with properties of the parton-medium interaction.

\section{Conclusions}

We have computed the variation of the suppression of back-to-back correlations for non-central 200 AGeV Au-Au collisions as a function of impact parameter $b$ for both in-plane and out-of-plane emission. Since the parameter determining the quenching power of the medium has been adjusted such that single hadron $R_{AA}$ for central collisions is described, the extension to back-to-back correlations, non-central collisions and the dependence on the angle with the reaction plane is computed without additional free parameters, given the underlying hydrodynamical evolution (which is in turn constrained by the need to describe bulk matter observables such as momentum spectra and elliptic flow $v_2$). Thus, the results represent a rather constrained prediction which can be used to test the validity of the combination of hydrodynamical evolution \cite{Hydro3d} with the radiative energy loss model \cite{Jet2,QuenchingWeights}.

The suppression pattern we find is not unexpected given previous results \cite{RP1,Correlations2}: As a function of associate hadron momentum, $I_{AA}$ is approximately flat but may exhibit a small rise. $I_{AA}$ increases with impact parameter, reflecting the reduced soft matter entropy production and a shortening of average in-medium pathlength. At the same time, a split between in-plane and out-of-plane emission grows, reflecting the spatial asymmetry in the initial state. There is an interesting similarity between the numerical values of $I_{AA}$ (and the magnitude of the split) and the values of $R_{AA}$ and the split between in-plane and out-of-plane emission observed there. There is no single physics reason which would dictate this outcome, rather one is looking at a cancellation of various different effects. An investigation in the geometry reveals that the physics underlying single hadron and dihadron observables in the model is rather different in terms of dominant point of origin, typical density being probed and average pathlength. A detailed comparison with data and computations for different system size, e.g. Cu-Cu may be helpful to investigate this further.

\begin{acknowledgments}

I'd like to thank K.~Eskola, J.~Ruppert and P.~Jacobs for helpful comments. The work of C.~Nonaka and S.~Bass on the hydrodynamical description of the medium is gratefully acknowledged. This work was financially supported by the Academy of Finland, Project 115262.
 
\end{acknowledgments}

\end{document}